\documentclass[12pt,preprint]{aastex}

\newcommand{\Msun}{{\ensuremath{\mathrm{M}_{\odot}}}\,}

\newcommand{\isofont}[1]{{\mathrm{#1}}}
\newcommand{\isomass}[1]{{\ensuremath{\isofont{^{#1}}}}}
\newcommand{\isocharge}[1]{{\ensuremath{\isofont{_{#1}}}}}
\newcommand{\isotope}[3]{{\ensuremath{\isocharge{#1}\isomass{#2}\isofont{#3}}}}

\newcommand{\I}[2]{{\isotope{}{#1}{#2}}}

\slugcomment{Version \today}

\begin{document}

\title{Nucleosynthesis in O-Ne-Mg Supernovae}
\author{R. D. Hoffman}
\affil{N Division, Lawrence Livermore National Laboratory, P. O.Box
  808, Livermore, CA 94550}
\email{rdhoffman@llnl.gov}
\and 
\author{B. M\"uller and H.-T. Janka}
\affil{Max-Planck-Institut f\"ur Astrophysik, Karl-Schwarzschild-Str. 1,
85741 Garching, Germany}
\email{bmueller@mpa-garching.mpg.de, thj@mpa-garching.mpg.de}

\begin{abstract}

We have studied detailed nucleosynthesis in the shocked surface layers of an 
Oxygen-Neon-Magnesium core
collapse supernova with an eye to determining if the conditions are suitable for $r-$process
nucleosynthesis. We find no such conditions in an unmodified model, but do find
overproduction of N=50 nuclei (previously seen in early neutron-rich neutrino winds)
in amounts that, if ejected, would pose serious problems for galactic chemical evolution.
\end{abstract}


\section{INTRODUCTION}

The site of the $r-$process has been the most
enduring mystery in nucleosynthesis theory since the publication of the 
seminal
papers in this field \citep{Cam57,Bur57}. 
Of particular promise (in their times and for some even today) have been the
many efforts suggesting Type II supernovae as the site with
the relevant conditions arising either in or near the exploding 
core (initially championed by B$^2$FH), with recent
attention focused on aspects of neutrino interactions 
\citep{wh92,woo94}, or in the outer layers \citep{fkt79,cow82,ech88}.
For reviews, see \cite{ctt91,wal97}.

The requisite conditions for $r-$process nucleosynthesis in
explosive scenarios with material freezing out from
nuclear statistical equilibrium (NSE) have been derived
with the general understanding that particular combinations
of three parameters, the entropy, electron fraction ($Y_e$), and expansion timescale, give
rise to specific features of the solar $r-$abundance pattern \citep{qw96}. 
As the wind evolves 
these parameters must sweep out a range of conditions that produce the 
many features of the solar $r-$process abundances, 
particularly the relative heights of the 2nd \& 3rd peaks \citep{woo94}.
Recent models of core
collapse still fall short of the necessary conditions to explain all
the abundance features of the solar $r-$process, especially the high mass
($A\ge 130$) component \citep{hof97}.

A site that has received recent interest involves explosions of
O-Ne-Mg cores in $8-10 \Msun$ supernovae. 
Successful SN explosions from such systems have been the subject of
much debate \citep{wch98}, some efforts found prompt explosions \citep{hnw84, nom84}, 
others did not \citep{bl85,bck87}, with the former now being completely ruled
out with the advent of modern treatments of neutrino transport. 
Appeals to late-time neutrino heating 
have also been suggested \citep{mw88}.
It should be noted that nearly every attempt has used a
common progenitor model \citep{nom84}.

Recent efforts to revive the idea of a low-$Y_e$, low-entropy scenario for $r-$processing
have included detailed nucleosynthesis
calculations \citep{wan05,nqm07}. In the former case, an unmodified SN model 
(computed without neutrino transport) provides a very low
explosion energy ($E_{exp}\sim 0.02$ B), with modestly neutron rich
conditions ($Y_{e,min} = 0.45$) and no $r-$process. To obtain it they artificially 
increased the shock
heating term to obtain larger explosions ($\sim 1$ B) and lower electron fractions 
$(0.14 \leq Y_{e,min} \leq 0.36)$ and suggest material experiencing these conditions 
must be ejected to explain the principle $r-$process features.

By contrast, \cite{nqm07} propose that the necessary
conditions arise in the shocked C-O layers above the O-Ne-Mg core at the location of
a very steep density gradient near the edge of the mass cut 
in these compact stars. This allows
for rapid shock wave passage accompanied by high peak
temperatures, giving rise to a short expansion timescale. 
Such a combination has been suggested as a viable $r-$process scenario
even at modest entropy with a neutron excess near zero \citep{jor04}.
As with the neutrino wind scenario, this production would be primary.
Ning et. al. appeal to this scenario to help bolster observational suggestions that
$r-$process nuclei with A $\geq$ 130 are
preferentially associated with a low mass SN component \citep{ish04,qw07}.

Although the initial conditions of \cite{nqm07} were taken
from a stellar evolution model \citep{nom84,nom87}, the 
evolution of temperature and density in the mass-shell trajectories was derived in a 
semi-analytical fashion assuming a shock speed in the region of interest approaching
$10^{10}$ cm s$^{-1}$. Using shock jump relations \citep{mm99},
they derived the entropy and expansion timescale and obtained 
resulting nucleosynthesis that did exhibit characteristics reminiscent
of $r-$process abundances. The authors conclude by imploring further research
to test their assertions in modern finely zoned supernova explosion simulations
of progenitors in the same mass range. 

Recently, \cite{kjh06} have calculated 
supernovae explosions in this mass range, considering an 8.8$\,M_\odot$
progenitor star with an O-Ne-Mg core \citep{nom84} and using
a sophisticated treatment of neutrino transport. The
nucleosynthesis studies presented here are based on an update
of these (spherically symmetric) simulations for a 
revised progenitor model, in which
the outer layers of the helium shell and a dilute 
hydrogen envelope were added (K.~Nomoto, private communication).
The new simulations also include an improved treatment of 
electron captures and inelastic neutrino scattering by nuclei in 
NSE \citep{lmp03, lmp07}.
The principle results of these simulations will be discussed in a 
separate publication \citep{mj08}. Some relevant information about 
the explosion dynamics can be found in \cite{jmkb07}.

\section{Nucleosynthesis in an exploding 8.8$\,$M$_{\odot}$ Star}

Our results survey explosive nucleosynthesis in 32 zones extracted 
from an 8.8 $\Msun$ SN model starting at the edge of the O-Ne-Mg
core and extending though the C-O and He layers.
The mass-cut that defined our inner most zone had an initial (pre-collapse)
radius of $7.717 \times 10^7$ cm with an enclosed mass of 1.363 $\Msun$.
The last zone studied had a radius at the onset of core-collapse of 
$1.1306 \times 10^8$ cm with an enclosed mass of 1.376 $\Msun$.
The amount of processed ejecta is 0.013 $\Msun$.
Exterior to this was a hydrogen envelope (70\% H, 30\% He) whose 
outermost zone at core-collapse was at 
radius $6.414\times 10^{13}$ cm with an enclosed mass of 2.626 $\Msun$.
The amount to total ejecta is the difference, 1.263 $\Msun$.

During shock wave passage, the temperature increased dramatically and
in most of the zones exceeded $T_9 = 7$. As such photo-disintegration will
disassemble any C or O into $\alpha-$particles and nucleons.
Each nucleosynthesis calculation started at a point in the expansion when
the temperature had declined to T$_9=9.0$ (or its maximum value if it 
never achieved this),
with the starting values of density and $Y_e$ also taken from the SN model,
and followed until the temperature declined to T$_9=0.1$ (or the minimum value 
achieved in the
SN model trajectory, in a few cases as high as T$_9=0.5$). Initial compositions 
for zones with an initial $Y_e \leq 0.5$ 
were assumed to be $\alpha$-particles and free neutrons, otherwise
free neutrons and protons. 
Since nucleosynthesis beyond the iron group requires particle capture
on seed nuclei that here must be built up by NSE, we define a simple
expansion timescale (in seconds) as the time for the temperature to decline from
T$_9$ = 5.0 (or its highest value attained) to 1/$e$ of this value.
See Table \ref{tab:nucl} for the initial conditions.

Our reaction networks are
described in \cite{pru05}. Neutrino induced reactions for free nucleons
and $\alpha-$particles were included in all calculations \citep{mcful95,mcful96}.
As reported by \cite{nqm07}, the inclusion of neutrino reactions was not crucial
to the final results of what is here essentially explosive nucleosynthesis.

\subsection{Results}

\begin{figure}
\begin{center}
\includegraphics[angle=270,width=\columnwidth]{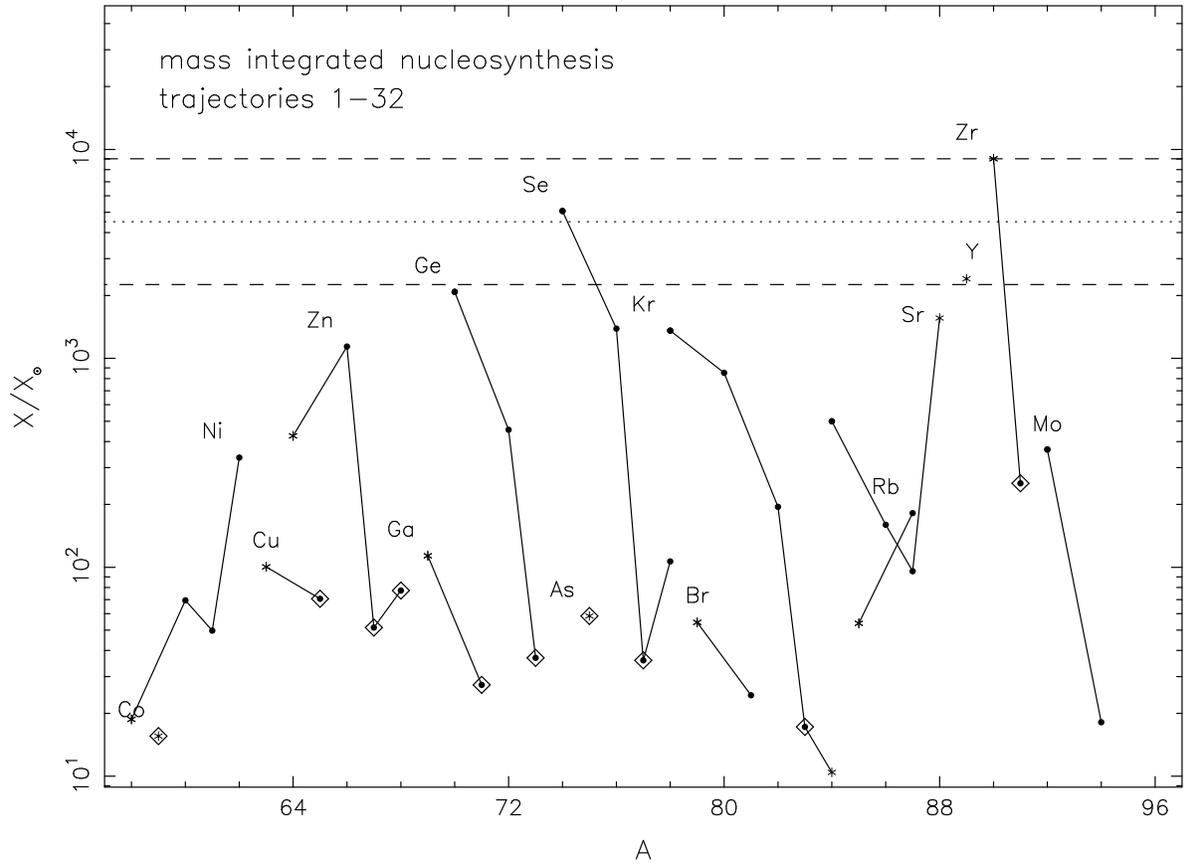}
\caption{
Mass weighted production factors characterizing the integrated nucleosynthesis in all
32 zones studied. Nuclides of a
given element are connected by solid lines, a diamond surrounding the
data point indicates production chiefly by a radioactive progenitor.
The horizontal lines represent a band of co-production for nuclei
made within a factor of two and four of \I{90}{Zr}.
\label{prodfac}}
\end{center}
\end{figure}

Figure \ref{prodfac} shows the mass weighted production factors normalized
to the total amount of ejecta given by

\begin{equation}
\label{proddef}
P(i)=\sum_j{M_j \over M^{\mathrm{ej}}}{X_j(i) \over X_{\odot,i}}.
\end{equation}

In this equation, the sum is over all 32 trajectories, $X_j(i)$ is the
mass fraction of nuclide $i$ in the $j$th trajectory, 
$X_{\odot,i}$ is
the mass fraction of nuclide $i$ in the sun \citep{lod03}, $M_j$ is
the mass of the $j$th trajectory,
and $M^{\mathrm{ej}}=1.263 \ \Msun$ is
the total mass ejected in the SN explosion whose energy was $\sim 0.1$ B
($\sim 0.2$ B in a corresponding 2D simulation).

We see no evidence of an $r-$process for the starting entropy and $Y_e$
conditions given in the model of M\"uller \& Janka (2007)
 principally because the expansion timescales 
of the trajectories containing the most mass are long ($\sim 0.1$ s)
and the entropies are small ($\sim 10-20$, see Table \ref{tab:nucl}).
By contrast the same quantities derived by \cite{nqm07} were
entropy = 139 $k_b$/nucleon,
and an expansion timescale to $e-$fold from T$_9 = 5.0$ of 0.013 sec.
The explosion energy agrees with model Q3 of \cite{wan05}, but
the $Y_e$ range is more neutron rich in the zones that dominated the 
nucleosynthesis.

We do note production of species above the iron group, but they terminate 
for nuclei in the N=50 closed shell. Attractive combinations of
expansion timescale and entropy (for the given $Y_e$) do appear to be
achieved for the trajectories farthest out, but these exhibit a
rapidly diminishing drop in the peak shock temperature and density
(the last two trajectories are in the extended hydrogen envelope, hence they have
$Y_e > 0.5$, they never reach NSE). 
In every calculation the free nucleons were all 
consumed in the build up of intermediate mass nuclei during the
alpha-rich freeze out. See Table \ref{tab:nucl} for the final $\alpha-$particle
mass fraction in each zone and the largest mass weighted production factor.

Indeed, the nucleosynthesis here is dominated by trajectories 7 to 12 which all
exhibit fairly long expansion timescales, entropies near 18, and 
initial values of $Y_e \leq 0.48$.
Between them they constitute 40\% of the total ejecta studied.
The production of the dominant N=50 isotopes (\I{90}{Zr} 
in particular) is most pronounced in zones 9 and 10. 
All of the co-produced species (those made within a factor of $\sim 4$ of the largest 
production factor) are made as themselves with the dominant reaction
flows occurring in the valley of stability.

This result is reminiscent of nucleosynthesis seen at early times in the neutrino-driven wind
of more traditional SNII \citep{wjt93,how93,woo94}, where an overproduction of N=50
isotopes presented problems for a model that at late times produced
a high entropy solution favorable to the $r-$process.
The problem here may not appear so striking, in that \I{90}{Zr} is also made with
\I{74}{Se}, but both are made at a very high production factor. 
Both \I{70}{Ge} and \I{90}{Zr} anchor two quasi-equilibrium clusters near
the iron group and N=50, their abundances   
determine the production of \I{74}{Se} and \I{92}{Mo}, respectively.
The other two light $p-$nuclei in between,
\I{78}{Kr} and \I{84}{Sr}, are not members of these quasi-equilibrium clusters
and owe their production to nuclear flows between them \citep{hof96}.
The dominant species (by mass fraction) for trajectory seven ($Y_e = 0.47$,
which made the most \I{92}{Mo})  
are \I{66}{Zn} and \I{60,62}{Ni} (all at 20\% by mass fraction),
and \I{90}{Zr}, \I{64}{Zn}, and \I{70}{Ge} (all $\sim 5$\%). 
As here, \cite{hof96} also saw
limited production of the heavier $p-$nuclei \I{94}{Mo}, \I{96}{Ru} and \I{98}{Ru},
which are made in equilibrium with \I{90}{Zr} but always at a level smaller than that
for \I{92}{Mo}. 
The N=50 nucleosynthesis can be eliminated in preference to light $p-$nuclei 
if $Y_e$ is constrained to be in 
a narrow range $(\sim 0.485)$. In the SN model of \cite{woo94},
the N=50 overproduction problem persisted if the
amount of material that experienced $Y_e \leq 0.47$ was greater than $10^{-4} \Msun$.
Here it is even worse (see below).

Interestingly, the recently developed $\nu p-$process does co-produce the
full range of light $p-$nuclei from \I{74}{Se} to \I{102}{Pd}, especially
\I{94}{Mo}, \I{96}{Ru} and \I{98}{Ru}, in quantities that
could for the first time explain their solar abundances
in early proton-rich neutrino wind models \citep{pru06,fro06}, but they fail to
co-produce \I{92}{Mo}. Details of the nuclear physics uncertainties
affecting the reaction flows that determine the solar ratio for \I{92}{Mo} and \I{94}{Mo} have
recently been explored \citep{fhp07}. 
Within current uncertainties in the proton-separation energies of \I{91-93}{Rh}, they suggest
this ratio could be achieved along with co-production of all
the light $p-$ nuclei between Sr-Pd in the $\nu p-$process if $S_p(^{93}{\rm Rh}) = 1.64 \pm 0.1$ MeV.

\subsection{Discussion}

If ejected, the values of the largest normalized production factors in 
these zones $(\sim 1000)$
would cause serious problems for Galactic Chemical evolution. 
Considering the largest production factor (\I{90}{Zr}), the enrichment 
by a single event relative to solar would be

\begin{equation}
{\left[ {{Zr}\over{H}} \right]}_{single} =
log \left( { {Y_{Zr}/M_{mix}}\over{X_{\odot}(Zr)} }\right)
\label{enrichment}
\end{equation}
where $Y_{Zr} = \sum_j X_j(Zr)\times M_j$ is the mass yield of \I{90}{Zr}
($1.75\times 10^{-4} M_{\odot}$ for the sum of zones 7-12 where \I{90}{Zr} is produced with
a mass fraction grater than $10^{-4}$),
$M_{mix}$ is the mass of ISM into which the ejecta of this single event mixes 
($\sim 3000 M_{\odot}$, \cite{tgjs98}),
and $X_{\odot}$ is the mass fraction of \I{90}{Zr} in the sun 
($1.53\times 10^{-8}$, \cite{lod03}). 
This is equivalent to the definition of the production factor (Eq. \ref{proddef}) 
with $M^{\mathrm{ej}}$ replaced by $M_{mix}$. For \I{90}{Zr}, the enrichment
is 3.8, {\it i.e.} nearly four times solar, implying a very rare event.

If $M_{mix}$ is typical of the mass that can be contaminated by such a SN and
assuming one occurred at most once at some average location over the
age of the Galaxy then the frequency $f_{SN}$ for it to occur is given from

\begin{equation}
\label{nexpected}
N_{expected} \sim { {f_{SN}M_{mix}t_{Gal}}\over{M_{gas}} } = 1
\end{equation}
where $M_{gas}\sim 10^{10} M_{\odot}$ is the mass of the gas in the Galaxy,
and $t_{Gal} \sim 10^{10}$ yr its age,  
giving a frequency of once every three thousand years. 
This is in
disagreement with recent best estimates that suggest explosions of 
O-Ne-Mg cores comprise 4\% of all core-collapse events \citep{poe07},
which translates (assuming a present day upper limit of
two SNII per century) to 0.0008 yr$^{-1}$. If one appeals to their firm upper limit of 20\%
(based on the many uncertainties in modeling the progenitors), then
this rate could rise to 0.004 yr$^{-1}$.

Over the history of the Galaxy, this enrichment is

\begin{equation}
{\left[ {{Zr}\over{H}} \right]}_{all} =
log {\left[
{
{{ Y_{Zr} f_{SN} t_{Gal}}\over{M_{gas} }} / X_{\odot}(Zr)
}
\right]}
\end{equation}
which gives 10-50 times the solar value.
We note that both
$M_{gas}$ and $f_{SN}$ may have been different in the past, but the ratio of the two
which is the rate of SN per unit mass of gas may vary much less as the SN progenitor formation
rate is proportional to the gas available.

\section{Conclusions}

We have studied in detail the nucleosynthesis in the shocked outer layers
of an O-Ne-Mg Type II SN to test the assertion that this might be a viable 
site for the $r-$process. While \cite{nqm07} derived shock conditions from 
the same progenitor model we use (\cite{nom84} plus the recent 
supplementation with a H-envelope) in a
semi-analytic manner with an {\em assumed} shock velocity),
our conditions arise from a detailed simulation \citep{mj08}.
We find conditions for nucleosynthesis
that are very different from those of \cite{nqm07}, especially in
regards to peak temperatures and expansion timescales. Consequently we
cannot support their assertion that this is a potential site for the 
$r-$process as we see neither the requisite conditions nor products. 
Our results are however very reminiscent of
previous calculations studying the nucleosynthesis in the neutrino
driven wind \citep{woo94,hof96}. 

The combination of low entropies ($\sim 20$) and electron fractions ($Y_e \leq 0.47$) 
in $5.5\times 10^{-3} M_{\odot}$
of ejecta from our model suggest that this one event would produce
nearly four times the solar abundance of \I{90}{Zr}.
This is unlikely, as observations of Zr are well established in metal poor stars where
enrichment to this level has not been seen. If this represents a very rare event, it
leaves little room for the production of Zr from other sources (such as the neutrino
winds of ordinary SNII), and still less for the $s-$process. But the initial mass function
suggests there are many stars with masses between 8-10 $M_{\odot}$, and recent studies indicate
that 4\% of these should end their lives as SNII \citep{poe07}. 
We are therefore forced to consider
that this is not a typical event (or even a very rare one) and that some aspect of the 
progenitor model, the explosion model, or the nuclear physics used in the determination 
of the nucleosynthesis is grossly in error.

On the later point, since the reaction flows producing the abundant species move 
along the valley of stability, we feel confident that the ingredients that went into
the calculation of the nuclear reaction cross sections
and the particle separation energies (so important in QSE)
are on firm ground \citep{Rau00}. 
There is no complication due to neutrino nucleosynthesis as the radii of
the zones considered, 100's of km, are too large for a substantial neutrino fluence.

On the stellar modeling side
our explosion model is based on a very modern treatment \citep{kjh06,mj08}, 
with issues of fallback being negligible. 
This leaves the pre-SN model, which was originally calculated as a He-core with 
a H-envelope subsequently added to it. There are numerous difficulties in accurately
calculating the pre-SN structure, including thermal pulses of the
unstable He shell source, mass loss, and dredge-up to name a few \citep{poe07}. 
We view this as the
most pertinent area where improvements should be made in understanding this important class
of stars and consider such improvements necessary to address the assertion that
they can serve as crucibles for half the species above iron.

We appreciate many valuable conversations with Yong-Zhong Qian
and S. E. Woosley. We also thank K. Nomoto for providing
us with his progenitor data and R. Buras, F. Kitaura, and A. Marek
for their inputs to the SN modeling project.

This work was performed under the auspices of the U.S. Department 
of Energy by Lawrence Livermore National Laboratory in part under 
Contract W-7405-Eng-48 and in part under Contract DE-AC52-07NA27344.
It was also supported, in part, by the SciDAC Program of
the US Department of Energy (DC-FC02-01ER41176).
The project in Garching
was supported by the Deutsche Forschungsgemeinschaft
through the Transregional Collaborative Research Centers SFB/TR~27
``Neutrinos and Beyond'' and SFB/TR~7 ``Gravitational Wave Astronomy'',
and the Cluster of Excellence EXC~153
``Origin and Structure of the Universe''. 
The SN simulations were performed on the national supercomputer
NEC SX-8 at the High Performance Computing Center Stuttgart (HLRS)
under grant number SuperN/12758.

\clearpage

\clearpage
\begin{deluxetable}{ccccccrccr}
\tabletypesize{\scriptsize}
\tablecaption{Outflow Characteristics
\label{tab:nucl}}
\tablewidth{0pt}
\tablehead{
  \colhead{traj.\tablenotemark{a}}
& \colhead{$r_{7}$}
& \colhead{T$_{9}$}
& \colhead{$s/k_b$}
& \colhead{$Y_e$}
& \colhead{$\tau_{\rm exp} (s)$}
& \colhead{Mass\tablenotemark{b}}
& \colhead{X($\alpha$)\tablenotemark{c}}
& \colhead{$^{A}{\rm Z}_{\rm P_{max}}$}
& \colhead{${\rm P_{max}}$}
}
\startdata
 1 & 1.08 & 9.2 &  30 &  .53 &  .046 &    5.00 &     .74 &  $^{45}$Sc &    2.95 \\
 2 & 1.09 & 9.5 &  29 &  .52 &  .051 &    5.00 &     .72 &  $^{45}$Sc &    1.98 \\
 3 & 1.23 & 9.0 &  27 &  .51 &  .058 &   10.00 &     .67 &  $^{49}$Ti &    3.44 \\
 4 & 1.32 & 9.1 &  25 &  .50 &  .064 &   10.00 &     .59 &  $^{62}$Ni &    2.04 \\
 5 & 1.47 & 9.1 &  24 &  .50 &  .073 &   10.00 &     .55 &  $^{62}$Ni &    7.05 \\
 6 & 2.14 & 9.0 &  21 &  .48 &  .092 &    5.00 &     .38 &  $^{64}$Zn &   45.53 \\
 7 & 2.27 & 9.1 &  20 &  .47 &  .100 &    5.00 &     .28 &  $^{74}$Se & 1001.58 \\
 8 & 2.40 & 9.0 &  19 &  .47 &  .107 &    5.00 &     .17 &  $^{74}$Se &  843.23 \\
 9 & 2.46 & 9.1 &  18 &  .46 &  .114 &   15.00 &     .12 &  $^{90}$Zr & 4928.74 \\
10 & 2.67 & 9.1 &  16 &  .45 &  .118 &   15.00 &     .04 &  $^{90}$Zr & 2589.07 \\
11 & 2.76 & 9.1 &  14 &  .46 &  .112 &    5.00 &     .10 &  $^{74}$Se &  629.45 \\
12 & 2.79 & 9.0 &  14 &  .47 &  .109 &   10.00 &     .14 &  $^{74}$Se & 1346.00 \\
13 & 2.70 & 9.0 &  12 &  .48 &  .104 &   10.00 &     .21 &  $^{64}$Zn &   60.97 \\
14 & 2.68 & 9.2 &  11 &  .49 &  .099 &   10.00 &     .22 &  $^{62}$Ni &   13.54 \\
15 & 2.50 & 9.0 &  10 &  .50 &  .092 &   10.00 &     .23 &  $^{62}$Ni &    4.68 \\
16 & 2.40 & 9.0 &  10 &  .50 &  .067 &    6.00 &     .24 &  $^{60}$Ni &     .82 \\
17 & 2.32 & 9.0 &  10 &  .50 &  .038 &    1.00 &     .31 &  $^{63}$Cu &     .29 \\
18 & 2.37 & 9.1 &  13 &  .50 &  .032 &    1.00 &     .43 &  $^{63}$Cu &     .43 \\
19 & 2.66 & 8.0 &  14 &  .50 &  .025 &    1.00 &     .49 &  $^{63}$Cu &     .49 \\
20 & 2.90 & 7.1 &  15 &  .50 &  .017 &     .10 &     .58 &  $^{63}$Cu &     .06 \\
21 & 2.98 & 7.0 &  16 &  .50 &  .015 &     .10 &     .60 &  $^{63}$Cu &     .06 \\
22 & 3.00 & 6.6 &  16 &  .50 &  .014 &     .05 &     .62 &  $^{63}$Cu &     .03 \\
23 & 3.07 & 6.3 &  17 &  .50 &  .014 &     .05 &     .64 &  $^{63}$Cu &     .03 \\
24 & 3.09 & 5.8 &  20 &  .50 &  .013 &     .05 &     .40 &  $^{63}$Cu &     .01 \\
25 & 3.32 & 4.9 &  42 &  .50 &  .013 &     .05 &     .00 &  $^{52}$Cr &     .00 \\
26 & 3.74 & 4.3 &  53 &  .50 &  .013 &     .05 &     .00 &  $^{40}$Ca &     .01 \\
27 & 4.77 & 3.2 &  88 &  .50 &  .013 &     .01 &     .02 &  $^{84}$Sr &     .05 \\
28 & 5.24 & 2.8 & 108 &  .50 &  .013 &     .01 &     .00 &  $^{41}$ K &     .00 \\
29 & 5.99 & 2.5 & 132 &  .50 &  .013 &     .01 &     .80 &  $^{39}$ K &     .00 \\
30 & 6.50 & 2.3 & 146 &  .50 &  .012 &     .01 &    1.00 &  $^{ 7}$Li &     .00 \\
31 & 7.21 & 2.0 & 185 &  .84 &  .010 &     .01 &     .33 &  $^{51}$ V &     .00 \\
32 & 9.07 & 1.4 & 381 &  .85 &  .006 &     .01 &     .30 &  $^{ 2}$ H &     .00 \\
\enddata
\tablenotetext{a}{Initial conditions at radius $r_7$ (in $10^7$ cm) when T$_9 \sim 9.0$ (or its peak value).}
\tablenotetext{b}{Mass of the zone in units of $10^{-4}M_\odot$. The mass interior to traj. 1 was 1.363 $\Msun$.}
\tablenotetext{c}{Final $\alpha-$particle mass fraction, the nucleus with the 
largest mass weighted production factor, and its value.}
\end{deluxetable}

\end{document}